# Energies and Entanglement in Multiply-coupled Phase Qubit Systems

**Z Thrailkill, S Kennerly, A Tyler and R C Ramos**

Low Temperature Laboratory, Department of Physics, Drexel University, 3141 Chestnut St., Philadelphia, PA 19104 USA

Email: zet23@drexel.edu

**Abstract**. The superconducting Josephson junction has been demonstrated to be a strong candidate for building quantum bits or "qubits" which are the components of a future quantum computer. In recent years, considerable theoretical and experimental effort have been focused on studying quantum properties of single qubits and two coupled solid-state qubits. We present results of numerical simulations of the energy spectra of more three phase qubits that are capacitively-coupled in different configurations. We discuss the ensuing entanglement between component qubits as manifested in avoided crossings and how these may play a role in building gates and transmitting qubit state information.

## 1. Introduction

An important requirement towards building a future solid-state quantum computer is the ability to couple multiple qubits together. Such systems of multiply-coupled qubits yield entangled states that are needed to implement key applications such as quantum state transfer, teleportation and error correction. For superconductor-based quantum computing, this has motivated both theoretical and experimental studies of two coupled superconducting charge qubits [1], phase qubits [2-3] and flux qubits [4]. There have also been studies of multi-particle entangled states [5], superconducting qubits coupled to a resonant cavity [6] and spectroscopic studies of three and four coupled flux qubits [7-8].

We present theoretical simulations of three coupled Josephson phase qubits that are capacitively-coupled to each other and arranged in two different configurations: a linear chain and a triangular network. Phase qubits are compact, tunable devices that have been recognized as one of the strongest candidates for quantum computing and has therefore attracted significant attention as to how they can be networked together [9-10]. We calculate the energy level spectrum of these two simple networks, describe how entangled states arise and briefly discuss quantum state transfer.

## 2. Numerical Simulations of Coupled Phase Qubits

To understand how the qubits interact with each other with different biasing currents, a simulation of the energy levels one would get when performing a spectroscopy experiment would be helpful. Using the RCSJ model, the Hamiltonian for a single Josephson phase qubit can be written as:

$$H = \frac{P^2}{2m} + W(\gamma) \quad \text{where } m = C(\frac{\Phi_0}{2\pi})^2 \text{ and } \Phi_0 = h/2e \tag{1}$$

$C$ is the junction capacitance, $P = C\dot{\gamma}$ is the canonical momentum and $W(\gamma)$ is the washboard potential for a current-biased Josephson junction. If we assume that several identical current-biased Josephson junctions are coupled to each other with identical capacitors of capacitance $C_C$, then the Hamiltonian for the entire system can be written as:

$$H = \frac{1}{2}\left(\frac{\Phi_0}{2\pi}\right)^{-2} [p_1, p_2, \cdots] \begin{bmatrix} & & \\ & C & \\ & & \end{bmatrix}^{-1} \begin{bmatrix} p_1 \\ p_2 \\ \vdots \end{bmatrix} + W_1(\gamma_1) + W_2(\gamma_2) + \cdots \quad (2)$$

The capacitance matrix $[C]$ is determined by the junction capacitances $C_J$, the coupling capacitances $C_C$ and the circuit topology [11]. A circuit can be represented by a graph where each vertex represents an individual junction and each connecting line represents a coupling capacitor. For example, two possible circuits are shown in Fig 1, together with their equivalent graphs.

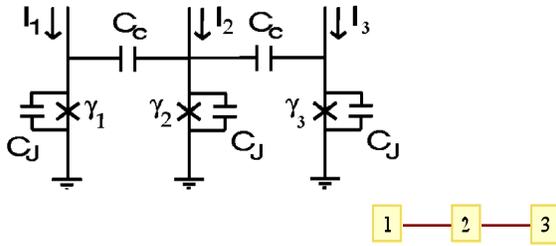
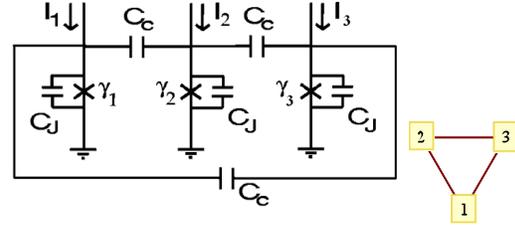

**Figure 1(a).** Series 3-Junction Circuit        **Figure 1(b)**. Triangular 3-Junction Circuit

The capacitance matrix can be calculated in a straightforward way. The **adjacency matrix** $[A]$ for a graph of n vertices is an $n \times n$ matrix with elements $A_{ij} = 1$ if vertices i and j are connected by a line and $A_{ij} = 0$ otherwise. The degree of a graph vertex is the number of lines connected to it (equivalently, the degree of vertex $n$ is the sum $\sum_j A_{nj} = \sum_j A_{jn}$). The **Laplacian matrix** $[L]$ for a circuit is formed by reversing the sign of its adjacency matrix and replacing each diagonal element $A_{jj}$ with the degree of vertex $j$. This can then be used to generate the capacitance matrix using the prescription $C_{ij} = C_J \delta_{ij} + C_C L_{ij}$. This procedure will be discussed in more detail elsewhere [12].

In the case of the three qubit system there are two possible configurations. The linear configuration has the same coupling between qubits 1 and 2, 2 and 3, but no direct coupling between 1 and 3. The other is the triangular configuration, which has the same coupling between all three qubits. The inverse capacitance matrix for the linear configuration and the triangular configuration are

$$\frac{1}{1+2\kappa}\begin{bmatrix} 1+\kappa-\kappa^2 & \kappa & \kappa^2 \\ \kappa & 1 & \kappa \\ \kappa^2 & \kappa & 1+\kappa-\kappa^2 \end{bmatrix} \qquad \frac{1}{1+2\kappa}\begin{bmatrix} 1 & \kappa & \kappa \\ \kappa & 1 & \kappa \\ \kappa & \kappa & 1 \end{bmatrix} \quad (3)$$

respectively, with $\kappa = C_c/(C_J + C_c)$. The symmetry in the triangular configuration results in an energy level degeneracy when all the biasing currents are equal.

To simulate systems of coupled-qubits one can *approximate* the washboard potential as a harmonic oscillator potential with a plasma frequency $w_p \propto \left(1-J^2\right)^{-1/4}$, and $J = I_b/I_c$ [2]. Without coupling, the eigenstates of the system are just the products of the individual qubit states $|n_1 n_2 n_3 \ldots\rangle = |n_1\rangle|n_2\rangle|n_3\rangle\ldots$ Ignoring any energy levels with more than one excitation, a basis can be formed using the direct product states to generate a Hamiltonian matrix where

$H_{ij} = \langle n_1 n_2 n_3 ... | H | n_1 n_2 n_3 ... \rangle$ Using the Jacobi transformation method for matrix diagonalization, one can determine the eigenvalues of the Hamiltonian for a given set of bias currents.

*2.2 Energy Level Spectrum of the linear 3-qubit system.*

Figure 2 shows the energy level spectrum of the first three excited states of the linear 3-qubit system when qubits 2 and 3 have fixed reduced bias currents J2 = 0.979 and J3 = 0.985, respectively and with a typical junction plasma frequency of about 5.5 Ghz. As seen from the two avoided crossings, ramping qubit 1's bias current J1 through these values of J2 and J3 maximally entangles qubit 1 with qubits 2 and 3, respectively. Furthermore, at the avoided crossings, the energy gap between the entangled states of qubit 1 and 2 is bigger than that between the entangled states of qubits 1 and 3. This is expected because qubits 1 and 2 are nearest neighbors (See Figure 1(a).) while 1 and 3 are farther apart so that the coupling between qubits 1 and 2 is stronger than that between qubits 1 and 3.

When J1 = J2 = J3, all qubits are maximally entangled with each other. When J1 is off resonance with J2 and J3, then the first qubit is not entangled but qubits 2 and 3 are.

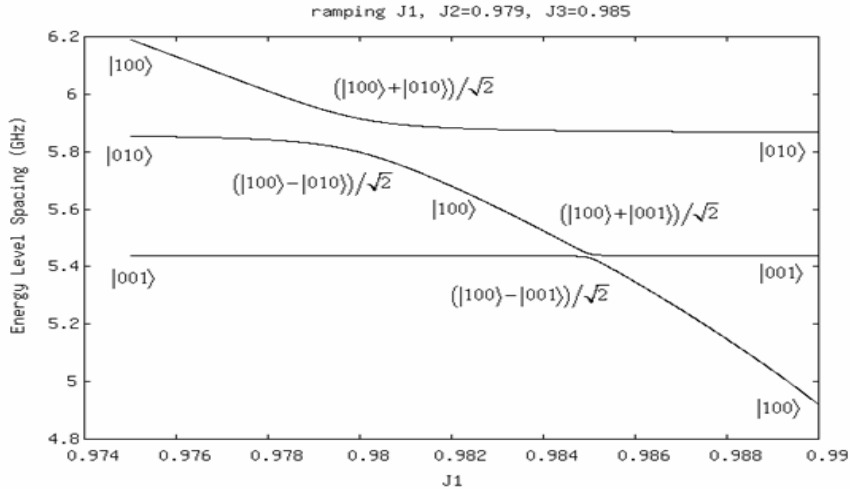

**Figure 2.** Energy Level Spectrum of the Series 3-Junction Circuit as a function of reduced current bias J1 when J2 = 0.979 and J3 = 0.985. Qubit 1 is maximally entangled with qubits 2 and 3 when J1 = J2 and J1 = J3, respectively, resulting in the two avoided crossings shown.

*2.1. Energy Level Spectrum of the Symmetric, Triangular 3-qubit system*

Figure 3 shows the energy level spectrum of the first three excited states for the triangle configuration of 3 qubits. When J1 is far from J2 and J3 the first qubit is not entangled but the second and third are. When J1 is equal to J2 and J3, there is a degeneracy caused by the symmetry of the system. Note that the $(|100\rangle - |001\rangle)/\sqrt{2}$ state is present no matter what J1 is. This is why that energy level remains constant. With J2 not equal to J3 it is possible to ramp J1 through J2 and J3 separately. This allows qubit 1 to be entangled with either qubit 2 or 3 separately. Although this is not shown here, the energy gap between entangled states of qubit 1 and 2 is equal to that between entangled states of qubits 1 and 3 because all the qubits are coupled equally to each other.

**3. Quantum State Transfer in Coupled Phase Qubit Systems**

Both theoretical and experimental work on the transfer of quantum state information from one qubit to another has recently attracted much attention [9-10]. In the present systems, this can be executed by tuning the biasing currents so as to dynamically entangle the qubit with the information with the qubit that is to receive the information. For example, suppose that the linear 3- qubit system is initially prepared so that qubit 1 is in the state $|1\rangle$ and qubits 2 and 3 are in the $|0\rangle$ state, and the system starts off with all three qubits unentangled, i.e. J1=0.983 (Fig 2). Here the system is in the $|100\rangle$ state, which is an energy eigenstate. A swap operation can be performed between qubits 1 and 2 by changing J1 to 0.979. Since the system is no longer in an energy eigenstate, it will evolve in time. After a time $\tau = \pi/\Delta E$ where $\Delta E$ is the energy gap between the two entangled states, it will be in the state $|010\rangle$. A similar operation can be done to transfer state information to qubit 3 by changing J1 to 0.985. Information transfer in the triangular configuration can be done in a similar fashion. However, the dynamics at the degeneracy point is different. This will be discussed in detail elsewhere [12].

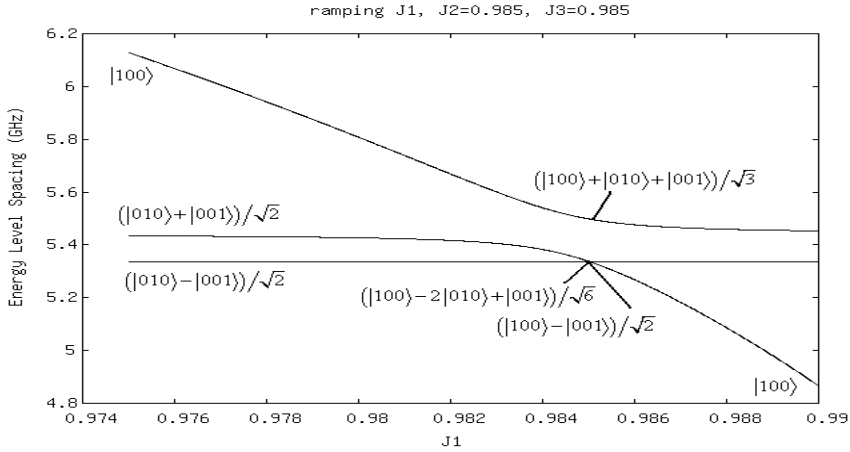

**Figure 3** Energy Level Spectrum of the Triangular 3-Junction Circuit as a function of reduced current bias J1 when J2 = J3 = 0.985. When J1 = J2 = J3, there is a degeneracy caused by the symmetry of the system.